\documentclass[doublecol]{epl2}
\usepackage{color}
\usepackage{graphicx}
\usepackage{dcolumn}
\usepackage{bm}
\usepackage{array}
\usepackage[table]{xcolor}
\usepackage{subfigure}
\newcommand{\PreserveBackslash}[1]{\let\temp=\\#1\let\\=\temp}

\newcolumntype{C}[1]{>{\PreserveBackslash\centering}p{#1}}
\newcolumntype{R}[1]{>{\PreserveBackslash\raggedleft}p{#1}}
\newcolumntype{L}[1]{>{\PreserveBackslash\raggedright}p{#1}}
\begin{document}

\title{Scaling behavior of online human activity}

\author{Zhi-Dan Zhao$^1$, Shi-Min Cai$^1$\thanks{\email{shimin.cai81@gmail.com}}, Junming Huang$^2$, Yan Fu $^1$, \and Tao Zhou$^1$}
\shortauthor{Zhi-Dan Zhao \etal}
\institute{$1$ Web Sciences Center, School of Computer Science and Engineering, University of Electronic Science and Technology of China, Chengdu 610054, P. R. China \\
$2$ Institute of Computing Technology, Chinese Academy of Sciences, Beijing, 100190, P. R. China}


\abstract {The rapid development of Internet technology enables human explore the web and record the traces of online activities. From the analysis of these large-scale data sets (i.e. traces), we can get insights about dynamic behavior of human activity. In this letter, the scaling behavior and complexity of human activity in the e-commerce, such as music, book, and movie rating, are comprehensively investigated by using detrended fluctuation analysis technique and multiscale entropy method. Firstly, the interevent time series of rating behaviors of these three type medias show the similar scaling property with exponents ranging from 0.53 to 0.58, which implies that the collective behaviors of rating media follow a process embodying self-similarity and long-range correlation. Meanwhile, by dividing the users into three groups based their activities (i.e., rating per unit time), we find that the scaling exponents of interevent time series in three groups are different. Hence, these results suggest the stronger long-range correlations exist in these collective behaviors. Furthermore, their information complexities vary from three groups. To explain the differences of the collective behaviors restricted to three groups, we study the dynamic behavior of human activity at individual level, and find that the dynamic behaviors of a few users have extremely small scaling exponents associating with long-range anticorrelations. By comparing with the interevent time distributions of four representative users, we can find that the bimodal distributions may bring the extraordinary scaling behaviors. These results of analyzing the online human activity in the e-commerce may not only provide insights to understand its dynamic behaviors but also be applied to acquire the potential economic interest.}

\maketitle

\section{Introduction}
The human behavior involving their daily activities is one of the highest complexity and complicated things because it is driven by countless unknown facts. Mining the human dynamics from these recorded large-scale data sets has become much more important for understanding their behavior patterns, and modeling the human dynamic behaviors helps us to explain many socioeconomic phenomena and find significant applications ranging from resource allocation, transportation control, epidemic prediction to personalized recommendation \cite{Barabasi2007,Zhou2008}. Thanks to the development of information technology, massive Internet data and resources make us easily realize the empirical analysis and modeling of human activity. One of the most attractive observation is the heavy-tailed nature of the interevent time distribution, which implies that the bursts of rapidly occurring events are separated by long periods of inactivity. Examples of empirical studies include the email communication \cite{Barabasi2005}, the surface mail communication \cite{Oliveira2005}, the cell-phone communication \cite{Candia2008}, the online activities \cite{Dezso2006,Zhou2008a,Goncalves2008,Radicchi2009,Zhao2012}, and so on. To understand the heavy-tailed phenomena of human dynamic behavior, the experts have put forward many mechanisms, such as the highest-priority-first queue model \cite{Barabasi2005}, the varying interest \cite{Han2008}, the memory effects \cite{Vazquez2007} and the human interacting \cite{Oliveira2009,Min2009,Wu2010} to mimic the temporal bursts.

On the other hand, the techniques of time series analysis are applied to investigate the evolutional data in real world. One of the most popular techniques is detrended fluctuation analysis (DFA) proposed by Peng \emph{et al}~\cite{Peng1994,Peng1993}, which can effectively quantify the long-range power-law correlations embedded in the nonstationary time series (or self-similarity process). It provides a simple scaling exponent $\alpha$ to represent the correlation characters of time series, and thus is applied to various research fields including heart rate dynamics~\cite{Peng1993,Peng1995,Ivanov1999,Ivanov2001,Ashkenazy2001}, financial time series~\cite{Liu1999,Gopikrishnan2000,Yamasaki2005,Ivanov2004,Jiang2009}, particle condensation~\cite{Tang2008}, Internet traffic~\cite{Cai2009}, musical rhythm spectra~\cite{Levitin2012}, etc. In recent,
Rybski \emph{et al}~\cite{Rybski2009,Rybski2010,Rybski2012} have applied the DFA to study the long-term correlations of the communication patterns (i.e., interactive activities) in online social networks, which is associated with the source of general Gibrat's law in economics. Meanwhile, Costa \emph{et al}.~\cite{Costa2002} recently proposed a technique, namely multiscale entropy~(MSE), to quantify the information complexity of physiologic time series over multiple scale. They further used the MSE to analyze the human heartbeat~\cite{Costa2005a}, which suggested that the time asymmetry is a fundamental property of healthy.
In the following works, the MSE was widely applied to analyze the EEG signals, which indicated that the
human brain variability increases with maturation~\cite{McIntosh2008ploscb} and the Alzheimer disease patients
usually had lower sample entropy on the small and medium time scales~\cite{Escudero2006physc}.
In the environmental field, Li and Zhang~\cite{Li2008SERRA} analyzed the long-term daily flow rates of the Mississippi River, and found that the sample entropy for flow rates generally monotonously increases with scale factor and the complexity
was beginning to decrease since 1940s. Moreover, based on the interevent time distributions and memory, Goh \emph{et al}~\cite{Goh2008} used the orthogonal measures to quantify the burstiness in many real interevent time series, and found that the origin of burstiness in human activity was much more correlated with the changes in the interevent time distributions.

The e-commerce is composed of the online business trades among humans and rating information based on the Internet Technology, in which the dynamic behaviors of human activity (i.e, trading or rating records) involve with a large number of useful knowledge for acquiring potential economic interest. In this letter, we first focus on the interevent intervals (i.e., time series) of human activity in e-commerce, and empirically investigate their scaling properties and information complexities both in the collective and individual levels based on the measures including the DFA, MSE and interevent time distribution. The rich results include that (1) The interevent time series of rating behaviors restricted to the types of medias show the similar scaling property that implies the collective behaviors of rating media follow a process embodying self-similarity and long-range correlation. (2) The different scaling exponents of interevent time series can be observed from the collective behaviors restricted to users' activities (i.e., rating per unit time), yet they both suggest the stronger long-range correlations existing in these collective behaviors. (3) The information complexities of collective behaviors are obviously distinguishable with the users' activities. (4) the extremely
small scaling exponents (indicating long-range anticorrelation) of a few representative individual users are mainly brought by the bimodal interevent time distributions.
\section{Materials}
The experimental data set is randomly sampled from Douban, which is a companion of e-commerce. It is similar to the Social Networking Services~$(SNS)$ that allows registered users to record information and create content related to movies, books, and music, yet it also can make a personalized recommendation for the registered users. We focus on users who perform more than 1000 rating actions on all three types of medias, which results in a set of 65 individuals. In the data set, we can find series of important history knowledge of registered users, such as user ID, item ID, rate, time stamp and item type, etc.
Note that the sample time resolution is second, and we here focus on the interevent time series defined as the intervals between two consecutive rating actions.

\section{Methods}
Herein we apply the DFA and MSE methods to
quantitatively understand the scaling behavior and
complexity of human activity in e-commerce. In order
to keep our description as self-contained as possible,
we should introduce the DFA and MSE methods briefly.
\section{Detrended Fluctuation Analysis}
We describe the process of DFA which involves
the following steps~\cite{Peng1994,Hu2001,Chen2002}.

(i) Starting with a time series $t(i)$, where $i \in [1,N]$ and N is the length of the series. We first integrate the series $t(i)$ and obtain $y(k) \equiv \sum\nolimits_{i = 1}^k {[t(i) - \left\langle t \right\rangle ]}$, where ${\left\langle t \right\rangle }$ is the mean. Meanwhile, $y(k)$ is divided into $N/l$ nonoverlapping boxes, each containing $l$ interevent intervals.

(ii) In the $k$-th box, we use a polynomial function ${y_l}(k)$ of order $n$
to represents the local trend. In the experiments, the order is selected as $n=2$,
and the algorithm is denoted as DFA-2.

(iii) We calculate the variance of residual time series after the detrending procedure,
\begin{equation}
F(l) \equiv \sqrt {\frac{1}{N}\sum\limits_{k = 1}^N {{{[y(k) - {y_l}(k)]}^2}} }
\end{equation}

(iv) Altering the box size $l$ and repeating the detrending procedure, we can obtain the
variances $F(l)$ as a function of box size $l$. A power-law relations between $F(l)$ and $l$ is
$F(l) \sim {l^\alpha }$. The $\alpha$ is a real value in the bounded range from $0$ to $1$,
where $\alpha>0.5$ means that the time series is correlated, $\alpha=0.5$ suggests that the time series is
same to the white noise (i.e., no correlation), and $\alpha<0.5$ indicates that the time series is anticorrelated.
\section{Multiscale Entropy}
We look back over the MSE method. The MSE is based on the simple observation that the complex signals generally exhibit the dynamics deviating far from perfect regularity and their multiscale complexity. The procedure of MSE is described as follows~\cite{Costa2002}:

(i) For a given time series, ${x_{1},x_{2},...,x_{N}}$, where N is the total number of time series.
we divide it into nonoverlapping boxes with the length $l$.

(ii) The averages of time series inside each boxes are deemed as the elements of a coarse-grained time series,
\begin{equation}
s_j^l = \frac{1}{l}\sum\limits_{i = (j - 1)l + 1}^{jl} {{x_i}} ,\begin{array}{*{20}{c}}
{}&{1 \le j \le N/l.}
\end{array}
\end{equation}
By altering the box length $l$, we can obtain many coarse-grained time series, which characterize the original
time series at multiple scales.

(iii) The sample entropy~\cite{Richman2000} is used to measure each coarse-grained time series. Thus,
we can find the relation between the entropy measure and scale factor (i.e., the box length $l$).

\label{Empirical results}
\section{Collective Level}
\begin{figure}
\begin{center}
\includegraphics[width=8cm]{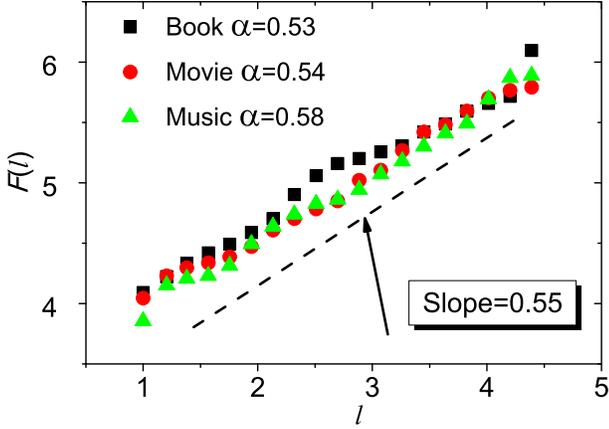}
\caption{\label{FIG1.MBM-DFA} (Color online) The results of
interevent time series measured by DFA. The symbols represent
the types of medias, book (black squares), movie (red circles),
and music (green triangles). The dash line is presented as guide eyes line.
The similar scaling behaviors suggest that the human activity in system follow
a in a long-correlated self-similar process.}
\end{center}
\end{figure}

The timestamp of data set is in precision of one second. Our focus is the interevent interval $\tau$
between consecutive actions, i.e. rating a media by a certain user in Douban.
The interevent time series are composed of these intervals in many ways.
From the view of whole system, we first investigate the interevent time series restricted to the types of medias.
Figure \ref{FIG1.MBM-DFA} shows that the results of interevent time series measured by DFA, in which we observe that
the scaling exponent $\alpha$ fluctuates in the small interval $[0.53,0.58]$. The values of $\alpha$ reveal similar scaling behaviors for all
types of medias and suggest that the interevent time series of rating media in the system evolves a process embodying self-similarity and long-range correlation. Furthermore, the scaling exponents slightly more than $0.5$ also imply the weak memory of the signals, which is consistent with the previous results found in other human activity~\cite{Goh2008}.

\begin{figure}
\begin{center}
\includegraphics[width=8cm]{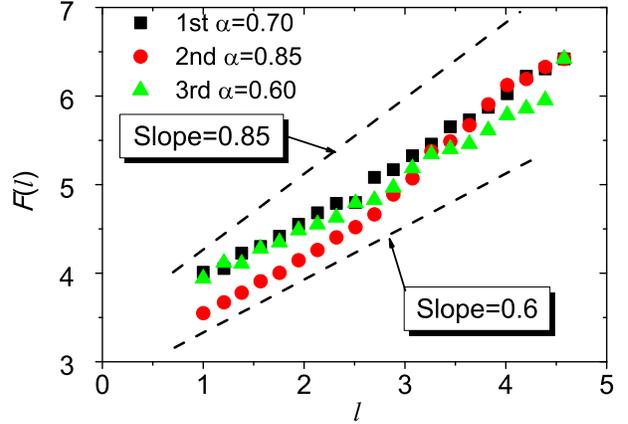}
\caption{\label{FIG2.Act-DFA} (Color online) The results of interevent time series measured by DFA. The interevent time series are constructed from the intervals between between consecutive actions in the three groups, $1st$ (black squares), $2nd$ (red circles), and $3rd$ (green triangles), respectively. The scaling exponents are 0.7, 0.85, and 0.65 corresponding to $1st$, $2nd$,
and $3rd$ groups respectively.}
\end{center}
\end{figure}

Although the human activity in whole system approximately obeys a common scaling law, we should
pay attention on the effect of individual user activity on their scaling
behaviors because the user activity is strongly associated with
the well understanding of human dynamics \cite{Zhou2008a,Zhao2012}. The
activity of an arbitrary user $i$, is defined as $A_i  = {{n_i }\mathord{\left/{\vphantom {{n_i } {T_i }}}\right.\kern-\nulldelimiterspace} {T_i }}$, where $n_{i}$ is the number of actions and
$T_{i}$ is the time difference between the first and last actions \cite{Ghoshal2006}. We sort these users in an increasing order
according to their activities, and then divide them into three groups which are indicated by low ($1st$), mid ($2nd$) and high ($3rd$) activity. The number of users in each group is 21 ($1st$), 22 ($2nd$), and 22 ($3rd$).
The interevent time series are constructed from the intervals between consecutive actions done by users in the three groups, respectively. Their results measured by DFA are shown in Fig. \ref{FIG2.Act-DFA}, which suggests that the scaling behaviors are different from the groups. For the $2nd$ group, the scaling exponent $\alpha=0.85$ indicates that the interevent time series opposes much stronger long-range correlations than these of other two groups. However, all the scaling exponents are much larger than $0.5$, which suggests that the long-range correlations generally exist in the interevent time series regardless of their activities.

\begin{table}
\caption {The scaling exponent $\alpha$ of original and shuffled time series measured by DFA.
Specifically, the first column corresponds to the scaling exponents~$\alpha$ of original time series,
and the second column indicates those of shuffled time series, respectively. Their difference is remarkable,
which verities that the memory effects generally exist in the online human activities.}
\centering
\begin{tabular*}{0.9\columnwidth}{@{\extracolsep{\fill}}ccc}
\hline \hline
       & $Orginal~\alpha$ & $Shuffled~\alpha$ \\
\hline
    Book   & \textbf{0.53} & 0.51 \\
    Movie   & \textbf{0.54} & 0.50 \\
    Music   & \textbf{0.58} & 0.50\\
    \hline
    $1$st   & \textbf{0.70} & 0.50 \\
    $2$nd   & \textbf{0.85} & 0.52 \\
    $3$rd   & \textbf{0.60} & 0.51 \\
\hline \hline
\end{tabular*}
\label{originalVSshuffled}
\end{table}

Additionally, to verify the the memory effects of online human activities don't arise from the pow law distributions of interevent time series, we reshuffle the original interevent time series to disturb their long-range correlations, and measure these shuffled ones by DFA method. In Tab.~\ref{originalVSshuffled}, we show the remarkable differences of scaling exponents between the original and shuffled time series, which suggest that the memory effects generally exist in online human activities.

\begin{figure}
\begin{center}
\includegraphics[width=8cm]{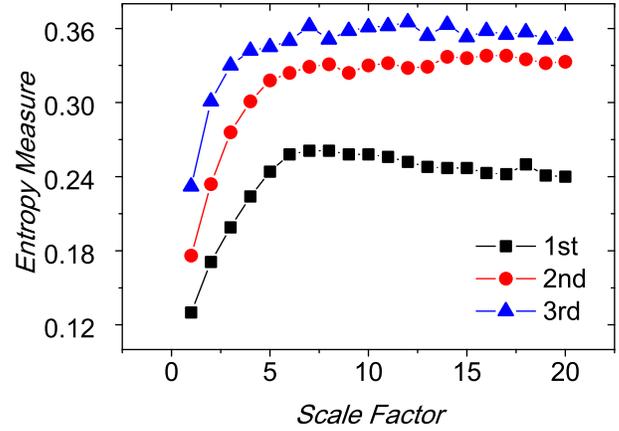}
\caption{\label{FIG3.Act-MSE} (Color online)  The results of interevent time series measured by MSE.
The sample entropy are obviously different from the activity of user behaviors, and much more information
is contained at large scale factors. The symbols indicate
$1st$ (black squares), $2nd$ (red circles) and $3rd$ ((blue triangles)) groups, respectively.}
\end{center}
\end{figure}

To further determine the differences among the three groups, we use the MSE method to quantify the information complexities of interevent time series. There are two guidelines, the higher information complexity is in correspondence with the larger sample entropy and the monotonic increase of the sample entropy indicates the much more information of interevent time series at large scale factors, to compare them. Figure \ref{FIG3.Act-MSE} shows that the sample entropy increases with the growth of user activity at all scale factors, and for all interevent time series they first increase at small scale factors and then approximately stabilize at the constant values, which suggests that the interevent time series constructed from the more active users become more complex and contain the much more information at large scale factors. These results demonstrate that the more actions done by users lead to the relative homogeneous interevent time series (i.e., the less extreme intervals). We should notice that although the information complexities are different from the the three groups, yet they don't directly associate with the degrees of long-range correlation.

\section{Individual Level}

\begin{figure}
\begin{center}
\includegraphics[width=8cm]{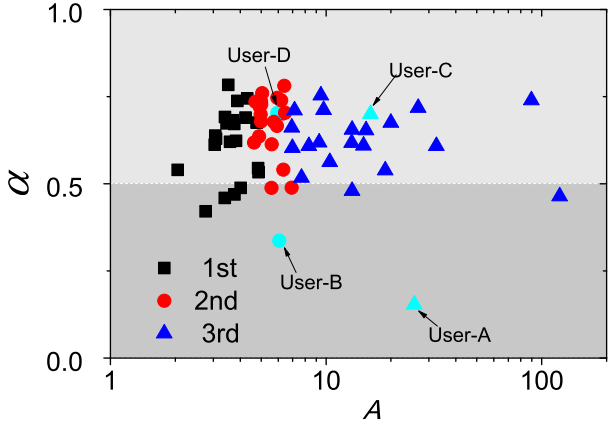}
\caption{\label{FIG4.Act-DFA-IND} (Color online) A scatter plot shows the detrended fluctuation analysis of the time series of different users' activities. Each point corresponds to a different user, indicating that there are significant differences between the scaling exponents and users' activity. Where ``black squares", ``red circles" and ``blue triangles" denote the individuals in $1$st, $2$nd and $3$rd activity group, respectively.}
\end{center}
\end{figure}

Although the obvious differences exist in the users behaviors at collective level, we still need to understand them and
explore the underlying mechanism of individual user behavior. Therefore, the interevent time series constructed from the total
65 individual user behaviors are further investigated by DFA. The users belonging to different
groups are denoted by the different symbols. In Fig. \ref{FIG4.Act-DFA-IND}, the scaling exponents are as a function of the activity of user, which shows that there is no direct correlation between them. Furthermore, most of scaling exponents greater than $0.5$ once demonstrate the existence of long-range correlations in the interevent time series of human activity in e-commerce. However, there are also abnormal user behaviors (e.g., users A and B seen in Fig. \ref{FIG4.Act-DFA-IND}) suggested by the scaling exponents much less than $0.5$.

This phenomenon urges us to observe and study these interevent time series constructed from the abnormal user behavior. Table \ref{singleactdfa} shows the basic statistical features of abnormal users A and B as well as these of two normal
users C and D. From the Tab. \ref{singleactdfa}, we can observe that the the similar activities may show the completely different scaling behaviors (e.g., users B and C) and the similar scaling behaviors don't means that they have same the activities (e.g., users C and D). We note that the Frequency denotes the event number of user behavior in Tab.1.

\begin{table}
\caption {The basic statistical features of the four selected typical users. The first column corresponds to the scaling exponents~$\alpha$, the second column indicates the activities, and the third column represents the frequency, respectively.}
\centering
\begin{tabular*}{0.8\columnwidth}{@{\extracolsep{\fill}}cccc}
\hline \hline
       & $\alpha$ & $Activity (day)$ & $Frequency$ \\
\hline
    User $A$  & 0.1534 & 25.68 & 3354\\
    User $B$  & 0.3369 & 6.07 & 3369 \\
    User $C$  & 0.7034 & 5.94 & 3557\\
    User $D$  & 0.7001 & 16.08 & 9725 \\
\hline \hline
\end{tabular*}
\label{singleactdfa}
\end{table}

\begin{figure}
\begin{center}
\includegraphics[width=8cm]{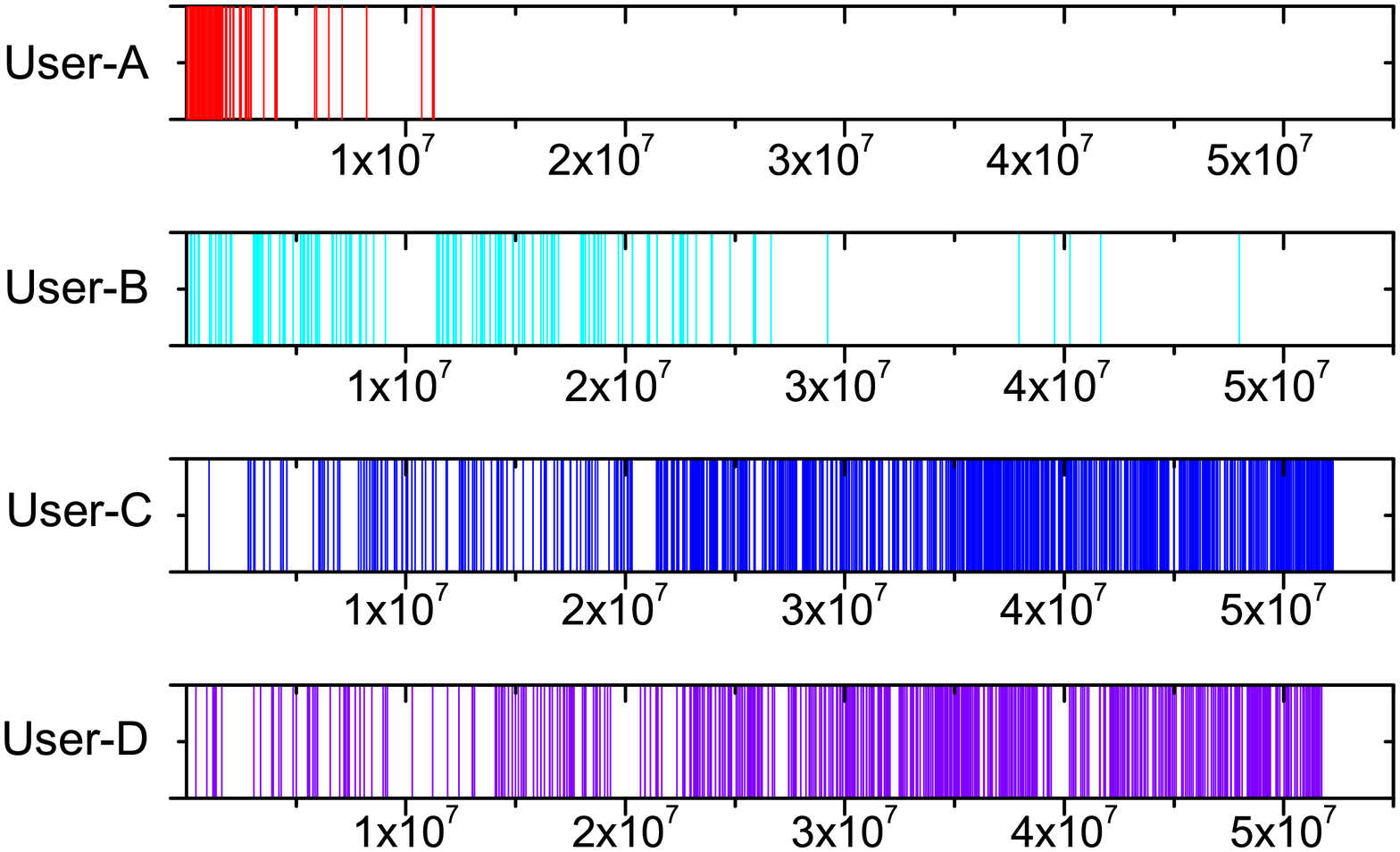}
\caption{\label{FIG5.Act-pattern} (Color online) The four online active patterns of users corresponding to Tab.~\ref{singleactdfa}. The horizontal axis denotes time, and each vertical line indicates an individual event.}
\end{center}
\end{figure}

To uncover the origin of the observed differences among scaling behaviors of individual users, we first present the rating activity evolving with time in Fig.~\ref{FIG5.Act-pattern}. We can straightly find that the active patterns of users $A$ and $B$ are very different from users $C$ and $D$. Concretely, the much more actions are occurred in users $A$ and $B$ at the initial stage, which results in the shorter time intervals, and the actions become much less or even absent (e.g, user A) when the time evolves.

\begin{figure}[htbp]
\subfigure[User-A]{
\begin{minipage}[t]{0.48\columnwidth}
\centering
\includegraphics[width=4.2cm]{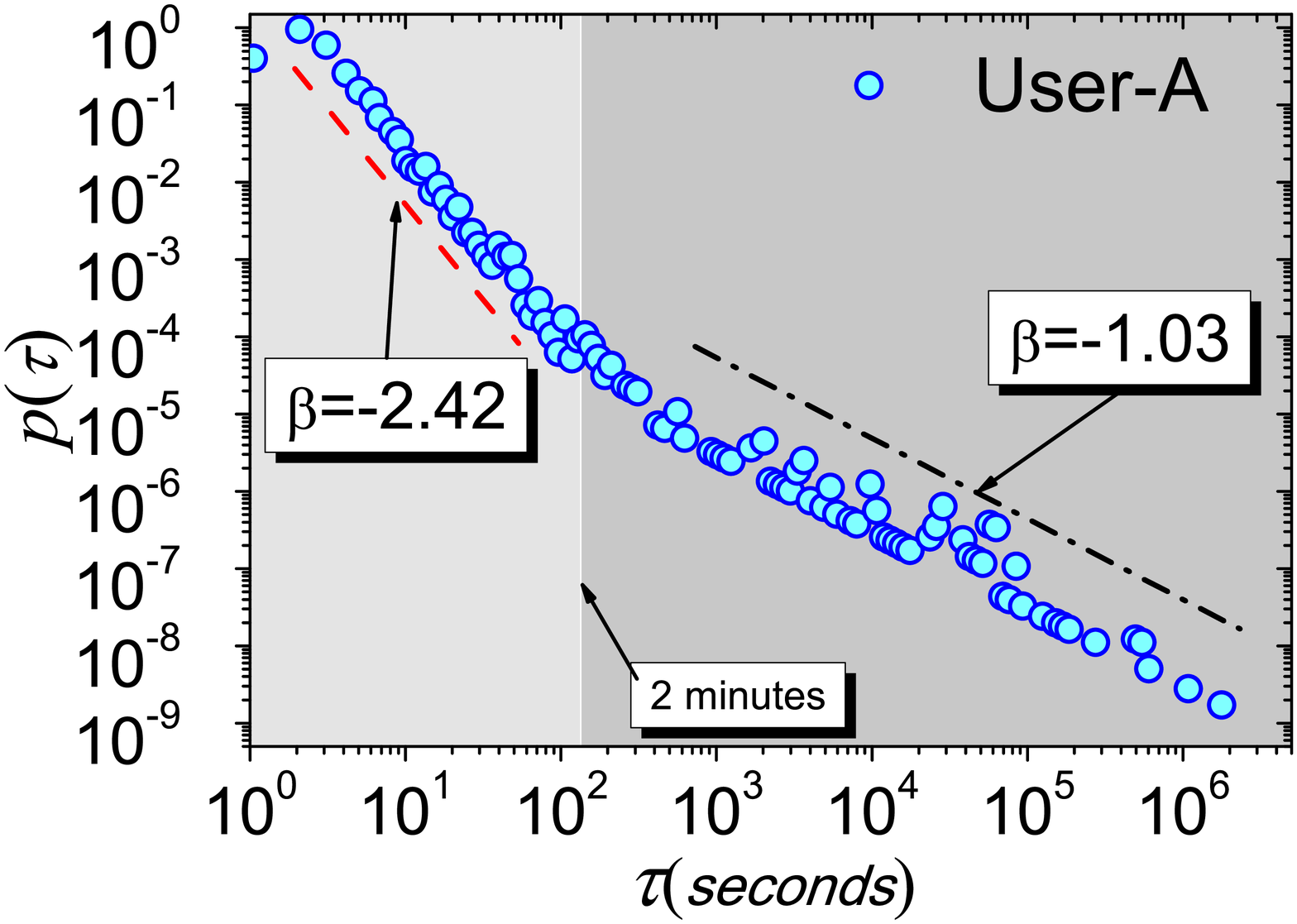}
\end{minipage}%
}
\subfigure[User-B]{
\begin{minipage}[t]{0.48\columnwidth}
\centering
\includegraphics[width=4.2cm]{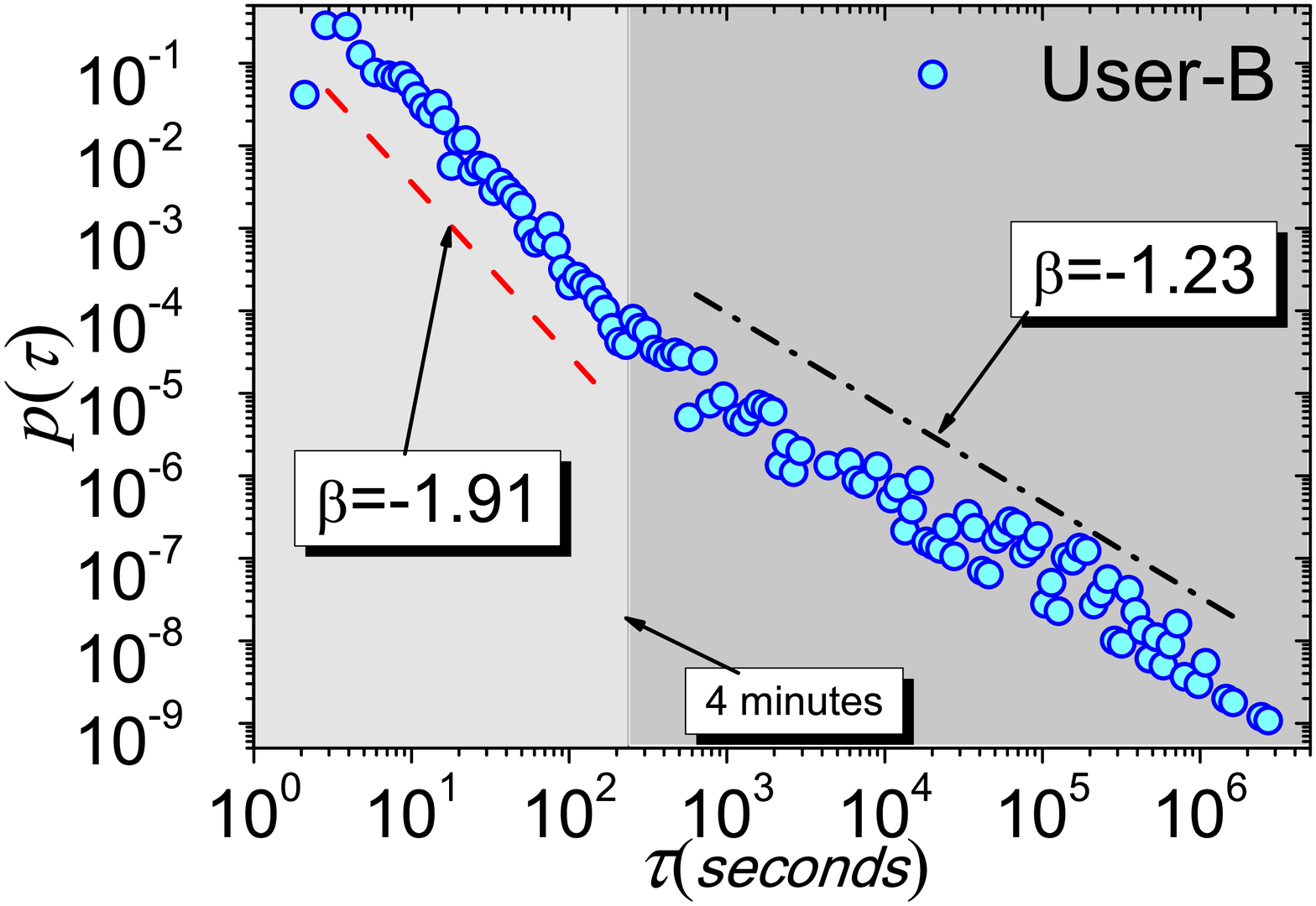}
\end{minipage}%
}\\
\subfigure[User-C]{
\begin{minipage}[t]{0.48\columnwidth}
\centering
\includegraphics[width=4.2cm]{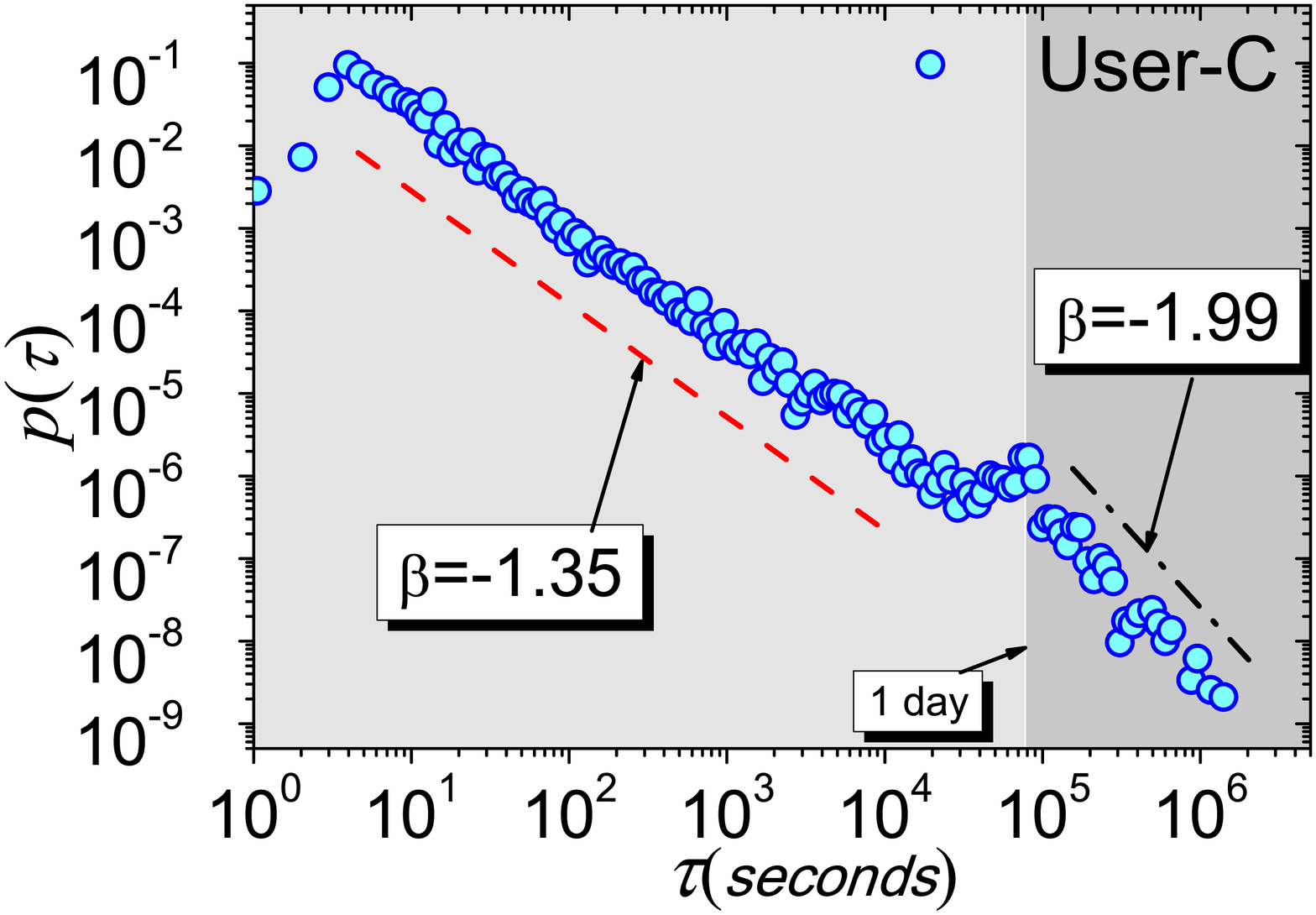}
\end{minipage}%
}
\subfigure[User-D]{
\begin{minipage}[t]{0.48\columnwidth}
\centering
\includegraphics[width=4.2cm]{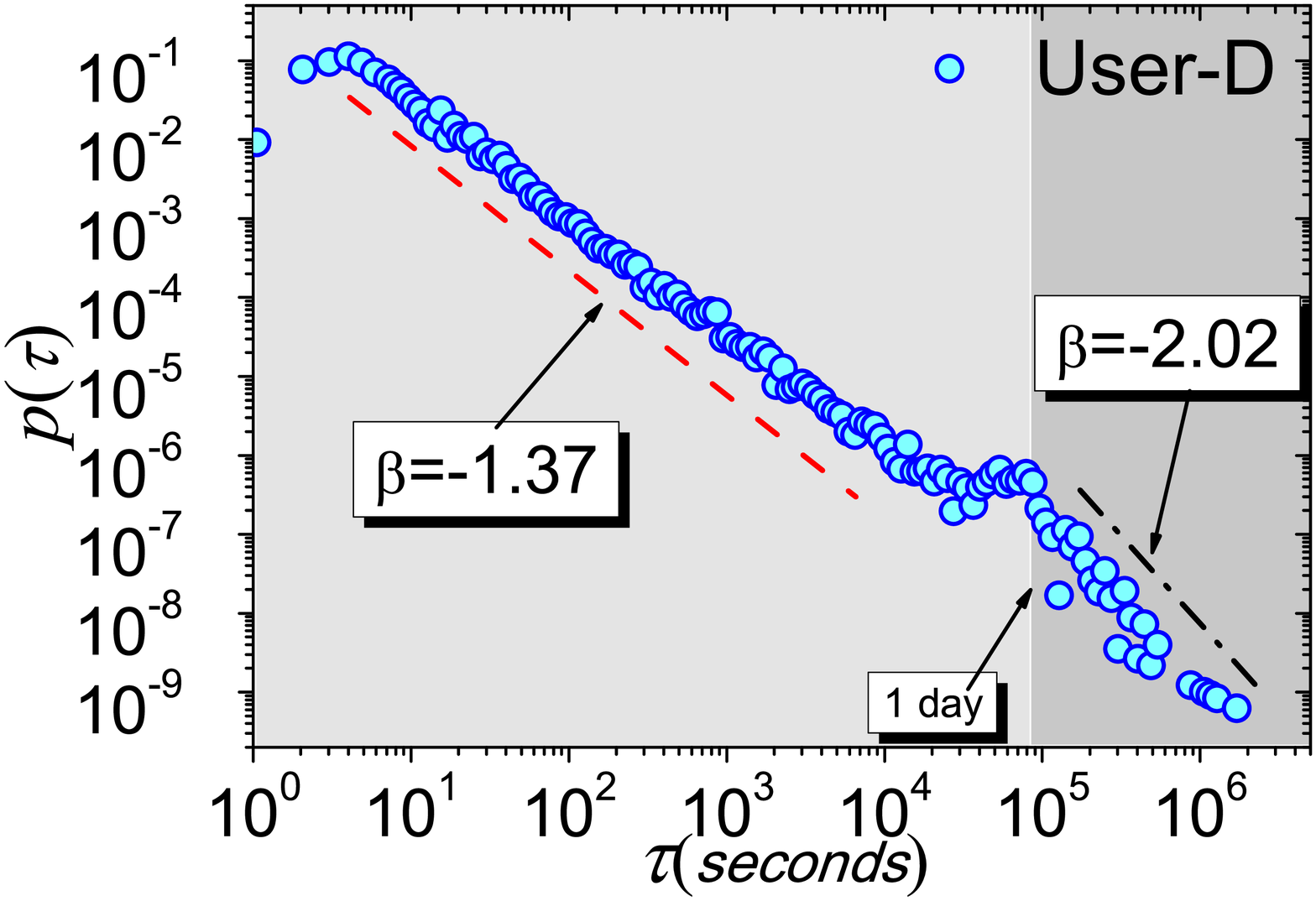}
\end{minipage}%
} \caption{Interevent time distributions.}
\label{intereventime}
\end{figure}

We propose the query on what is the specific active pattern for four users, and therefore statistically illustrates the interevent time distributions of the four users in Fig.~\ref{intereventime}, where the interevent time series are defined as the time intervals between consecutive actions by a certain user.
The results confirm that the interevent time distributions of these online human activities follow a power law form, $p(\tau)\sim\tau^{-\beta}$. However, we should note that there is a cutoff at the minute scale for the
the interevent time distribution of users A and B, respectively, while for those of users C and D,
the blurry cutoffs are at the day scale. The scales of minute and day are a typical decay length of online interests,
for example, user actions usually appear within a day in the microblogging systems~\cite{Zhao2012}.
Based on the phenomena of bimodal interevent time distributions, it can be found that the power-law exponents $\beta$ of users A and B change from big to small, yet these cases are quite reverse for users C and D. These changing trends of
the bimodal interevent time distributions vividly describe the differences among active patterns of individual user behavior,
which suggests that the short and long interevent intervals for users A and B alternately occur
and the short (or long) interevent intervals for users C and D continuously emerge.
Thus, we think that the scaling behaviors strongly associate with these changing trends of interevent time distributions for human activity in e-commerce although the potential dynamic mechanisms of online individual activity are similar.

\section{Conclusions}
We conclude that our empirically analysis, including the scaling behaviors and information complexities of human activity (i.e., rating the medias including music, book, and movie) comprehensively investigated by using DFA and MSE methods, provides the well understanding of behavior patterns of human activity in e-commerce. We also find that, for all rating behaviors corresponding to the types of medias, they display the similar scaling property with exponents ranging from 0.53 to 0.58, which implies that the collective behavior pattern of rating media follows a process embodying self-similarity and long-range correlation. Furthermore, by dividing the users into three groups based on their activity, we observe that the scaling exponents among three groups are different, yet they both suggest the stronger long-range correlations exist in the collective behaviors. Meanwhile, the information complexities of human activity quantified by MSE confirm the differences of scaling behaviors in these three groups. Moreover, we study the behavior patterns of human activity at individual level, and find that
the behaviors of a few users have extremely small scaling exponents associating with long-range anticorrelations.
By comparing with the distributions of interevent time of four representative users, we think that the different scaling behaviors are brought by the bimodal forms of the interevent time distributions.

\section{Acknowledgments}
We thank Ming Tang for the valuable discussion. This work is jointly supported by the NNSFC(Grant Nos.90924011, 60933005, 61004102, 11105025), China Postdoctoral Science Foundation (Grant No. 20110491705), the Specialized Research Fund for the Doctoral Program of Higher Education (Grant No. 20110185120021). ZDZ appreciates the financial support of the Fundamental Research Funds for the Central Universities (Grant No. ZYGX2011YB024).

\end{document}